# Atmospheric oxygen in Mn doped GaAs/GaAs(0 0 1) thin films grown by molecular beam epitaxy

J.F. Xu[a], P.M. Thibado[a], C. Awo-Affouda[b], R. Moore[b], V.P. LaBella[b]

[a]Department of Physics, University of Arkansas, Fayetteville, AR 72701, USA
[b]College of Nanoscale Science and Engineering, University at Albany-SUNY, Albany, NY 12203, USA

Abstract
Mn doped GaAs thin films were grown using molecular beam epitaxy at high and low substrate temperatures. The elemental concentration depth profiles in the thin films were determined by using Auger electron spectroscopy combined with ion etching. The Mn concentration is higher near the surface and then decreases with depth for films grown at high substrate temperatures. The Mn concentration profile is much more uniform when films are grown using a low substrate temperature. What was unexpectedly found are high levels of oxygen in the low substrate temperature grown thin films.

1. Introduction
Since the discovery of the ferromagnetic semiconductor of Mn-doped GaAs (GaMnAs) by Ohno et al. [1], it has been the subject of intense experimental [2,3] and theoretical work [4,5]. The combination of ferromagnetism with the versatile semiconducting properties makes it promising for future spintronics applications [6,7]. GaMnAs can be described qualitatively as a random alloy in which Mn substitutes at Ga sites and takes the dual role of acceptor and local magnetic moment [8,9]. Unfortunately high concentrations ($\sim 10^{21}$ cm$^{-3}$) of substitutional Mn are required for ferromagnetism to occur [1]. Worse still Mn has a low solubility limit in GaAs ($\sim 10^{19}$ cm$^{-3}$) [10]. At present, some of the best samples of GaMnAs have been grown using molecular beam epitaxy (MBE) and using low substrate temperatures [11]. The low-temperature MBE allows one to dope GaAs with Mn over its solubility limit, making it possible to realize a III–V-based diluted magnetic semiconductor (DMS) [1]. The low substrate temperature is also thought to be necessary to mitigate the formation of the thermodynamically more stable second phases such as MnAs [10]. The second phases can, in turn, result in the formation of As antisites (As$_{Ga}$) and Mn interstitial (Mn$_I$) defects, both of which compensate substitutional Mn acceptors (Mn$_{Ga}$) [12]. Because the ferromagnetism in a DMS structure is hole mediated, a reduction in the hole concentration due to these compensating defects leads to a suppression of the ferromagnetic Curie temperature [13]. Edmonds et al. [14] have grown GaMnAs films at low substrate temperatures (~200 C) and found Curie temperatures increased 60–70K after annealing films at 175 C. Recent work also indicated Curie temperatures up to 150K in annealed GaMnAs epilayers grown with a low substrate temperature of 250 C (annealing temperature also at 250 C) [15]. They pointed out that annealing likely decreases the concentration of Mn$_I$, which would enhance the hole density in the sample. Furthermore, they noticed the sample thickness limited the maximum attainable Curie temperature in the epilayers, which revealed that the free surface of GaMnAs may play an important role in the ferromagnetic properties.

It is interesting to note the highest Curie temperature around 150K achieved for this system is well below the 300K predicted by theory [16]. There would therefore still seem to be considerable potential for improving on the quality of the GaMnAs films produced so far. Consequently, precise control of the sample growth is critical and important. What are needed are investigations of the Mn doping properties, including distribution, diffusion and segregation as a function of the growth parameters. In this paper, we present the elemental depth profiles in GaMnAs grown at high and low substrate temperatures.

2. Experimental details
Samples were prepared in an ultrahigh vacuum ($\sim 2 \times 10^{-10}$ Torr) MBE growth chamber (Riber 32) which includes Ga and Mn effusion cells together with a two-zone As valved-cracker cell. The MBE chamber is also equipped with a reflection high-energy electron diffraction (RHEED) system. Commercially available, ''epi-ready,'' semi-insulating 2 inch diameter GaAs(0 0 1) +/- 0.1 degree wafers were cleaved into quarters. One quarter was mounted on a 2 inch diameter standard MBE molybdenum block using indium as solder. The substrate was heated to 580 C while exposing the surface to As$_4$ to remove the surface oxide layer. Afterwards, a thin buffer layer of GaAs was grown on the substrate for 1–5 min. During this time RHEED oscillations were used to determine that the growth rate of the GaAs was 780 nm/h. Next, the substrate temperature was set to the desired growth temperature of either 580 or 250 C. GaMnAs films were then grown for 1 h. After growth, the sample was cooled down and removed from UHV chamber.

The elemental composition profiles in the GaMnAs films were determined by using Auger electron spectroscopy (AES) combined with ion etching. The films were sputtered with an argon ion gun using a focused spot size of about 1mm in diameter. The depths of the sputtered holes were determined by measuring the crater depth using a Tencor stylus profilometer. In order to accurately calibrate the sputter rate, a standard sample had to be prepared. For this a semi-insulating GaAs substrate was implanted with Mn. Using this implanted sample as a calibration standard, the relative sputter rate was determined. This calibration was used to correct the AES depth profiles obtained from the MBE grown samples.

3. Results

AES studies on films after growth show the presence of significant concentration O and Mn on the surface of GaMnAs films. A typical AES spectrum for a film grown at substrate temperature of 580 C is shown in Fig. 1. By ion etching the surface of the film layer by layer, the AES depth profiles were obtained.

The elemental concentration (atomic percent) depth profiles in a GaMnAs thin film grown at a high substrate temperature (580 C) is shown in Fig. 2(a). Four elements, As, Ga, Mn, and O were found in the film. The percentage of As remains basically unchanged around 50% for all depths measured. The percentage of O drops to a low level near the surface and then stays at zero. The percentage of Mn is highest at the surface (14%), then decreases to zero at some depth around 800nm in the sample. This depth is the expected thickness of the thin film deposited on the substrate. The percentage of Ga is only 31% at beginning and it increases steadily with depth until it reaches 50% around 800 nm. What is most interesting is that the Mn and Ga concentrations change together with Mn going down, Ga going up, and the sum remaining constant near 50%.

When the thin film is grown using a low substrate temperature (250 C), the Auger elemental profile results are shown in Fig. 2(b). All the elemental profiles are flat until a step at about 800 nm, which is the film–substrate interface. For this sample, we also observed four elements: As, Ga, O and Mn in the film, but the elemental percents and depth profiles are significantly different from those in Fig. 2(a). First, the atomic percent of As in the thin film is around 43%, which is lower than that in Fig. 2(a). The Ga concentration keeps a constant value of 34% in the entire epilayer. Once the GaAs substrate is reached, the atomic percents for Ga and As reach 50%, as expected. The Mn concentration is about 5%. The primary finding is that the percentage of O in the thin film is about 18% and it stays constant all the way to the substrate. This is surprising because our MBE chamber does not have an intentional source of oxygen inside.

4. Discussion

The Mn concentration profile for the sample grown at high substrate temperatures is unusual (see Fig. 2(a)). It starts at zero near the substrate interface (about 800nm) and grows linearly to about 14% at the film's surface. This happened even though the Mn atoms were deposited uniformly throughout the thin film. If we calculate the area under the Mn concentration profile and divide this by the thickness of the same, we find the average Mn concentration would be about 7%. This number is close to what we expected to achieve. It must be that some of the Mn deposited in the earlier planes is not staying in that plane but migrating up as the film grows thicker. The phenomenon of floating along the growth front has been observed during the growth of GaInAs and MnGe [17,18]. Usually, this happens because the surface free energy is lower for one atom type, so the system prefers to have that atom's bonds broken at the surface compared to the other atoms present. On the other hand, at the lower substrate temperature we see a very uniform Mn concentration profile (see Fig. 2(b)). Given that the Mn wants to float along the growth front at high substrate temperatures, this ability must be frozen out for low substrate temperatures.

It is useful to analyze the various atomic percentages when the sample is grown at high substrate temperatures (see Fig. 2(a)). This sample has a small amount of oxygen near the surface and then very little inside the sample. In addition, the As percentage stays basically constant around 50% throughout the thin film. The interesting behavior is between the Ga and Mn atomic percentages. As the Mn concentration slowly grows from the substrate to the films surface, the Ga concentration moves in concert. The two levels are clearly tied together. As more Mn is incorporated, less Ga is incorporated. The whole time As level stays at 50%. This is a strong evidence that the Mn atoms occupy Ga atomic sites. Furthermore, since the Ga concentration plus the Mn concentration adds to about 50%, it is unlikely that Mn goes into interstitial sites or onto As sites.

After analyzing all the atomic percentages for the sample grown at a low substrate temperature (see Fig. 2(b)), the biggest surprise is the large percentage of oxygen in the thin film. Equally surprising is that the oxygen concentration persists throughout the entire thin film but there is none in the substrate. Given that there is no oxygen in our MBE growth chamber and the sample grown at 580 C has no oxygen inside, we believe this oxygen must come from the air after the sample was removed from the vacuum chamber. Previous studies of low-temperature grown GaAs have shown that a large number of defects exist in these films [19]. This results in a much lower atomic density in these films (e.g., ~80–90%), when compared to those grown at the normal high substrate temperature. We believe, due to a lower density, these films must be more porous and oxygen from the air must be able to easily diffuse into the thin film right down to the substrate.

Another interesting feature related to the atomic percentages in the low temperature sample (see Fig. 2(b)) is that the As concentration is just above 40% throughout the thin film and not 50%. Since oxygen was added to the sample after growth this would shift all the atomic percentages down in a linear way. If we subtract off the oxygen level of about 18% and then scale the remaining three back to 100% we find these levels: As = 52%, Ga = 42%, and the Mn = 6%. Thus, similar to the high temperature grown material, As is close to 50%. The Mn and Ga are also correlated, making it likely that Mn is replacing Ga in the sample. Of course, the film does have a lot of defects so it is much more difficult to draw these conclusions with certainty.

By looking at the various bond energies between O and the other elements, one can make a guess about which element the oxygen is attracted to most. We know that Mn–O and As–O compounds have much larger bond energies compared to Ga–O [20], so they would be favored. However, theoretical calculations have also revealed that the formation energy for As–O is greater than 6.0 eV because of the very large size mismatch between As and O [21]. From this it is reasonable to think that Mn–O bonds are primarily forming in the film.

It is unclear what the role of oxygen in the films will be. A recent study [22] showed that a significant fraction of Mn ions in GaMnAs occupies nonsubstitutional, interstitial sites ($Mn_I$). It is thought that oxygen in the sample may bond with these interstitial atoms. A decrease of number of Mn interstitials is thought to be favorable for ferromagnetism in GaMnAs. Therefore, surprisingly, it may be beneficial to have oxygen in these films when considering ferromagnetism.

5. Conclusions
In conclusion, GaMnAs films were grown using MBE at high and low substrate temperatures. For the sample grown at a high substrate temperature, Mn atoms seem to float along the growth front giving a graded concentration profile. Samples grown at low substrate temperature give a uniform distribution for the Mn profile, however high levels of oxygen are present in these films after exposure to air. The correlated percentage changes between Ga and Mn gives a strong indication that Mn atoms occupy Ga lattice sites.

Acknowledgment
The authors would like to acknowledge the support for this work from National Science Foundation under grant number DMR–0405036, DMR-Career-0349108, and Marco Interconnect Focus Center.

Fig. 1. Auger electron spectrum taken from the surface of a GaMnAs film grown using a Mn cell temperature ($T_{Mn}$) of 900 C and substrate temperature $T_s$ of 580 C.

Fig. 2. Elemental percentage depth profiles of Ga, As, Mn and O in GaMnAs films grown using a Mn cell temperature $T_{Mn}$ of 900 C but different substrate temperatures $T_s$: (a) $T_s$ = 580 C and (b) $T_s$ = 250 C.

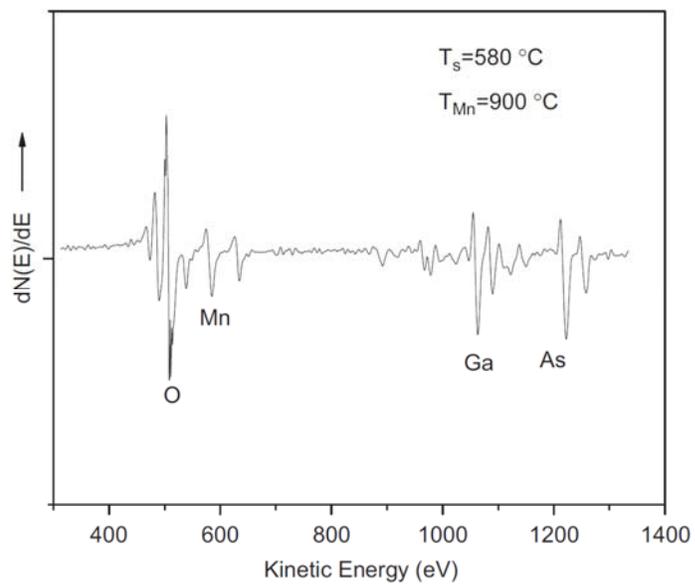

Figure 1.

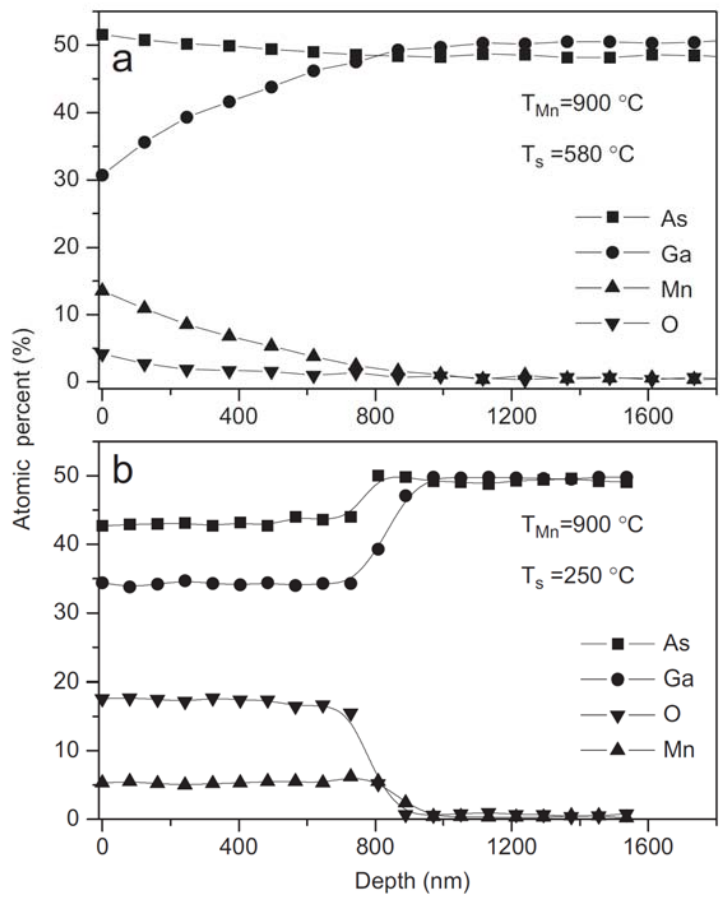

Figure 2.